\documentclass[12pt,a4paper]{article}
\usepackage{mathrsfs}
\usepackage{epsfig}
\usepackage{slashed}
\usepackage{graphicx}
\usepackage{subfig}
\usepackage{amssymb}
\usepackage{amsmath}
\usepackage{amsthm}
\usepackage{multirow}
\begin{document}

\title{Probing Anomalous FCNC Top-Higgs Yukawa Couplings at the Large Hadron Electron Collider}
\author{
        Wei Liu$^{1}$,
        Hao Sun$^{1,2}$\footnote{Corresponding author: haosun@mail.ustc.edu.cn \hspace{0.2cm} haosun@dlut.edu.cn},
        XiaoJuan Wang$^{1}$,
        Xuan Luo$^{1}$\\
{\small $^{1}$ Institute of Theoretical Physics, School of Physics $\&$ Optoelectronic Technology,} \\
{\small Dalian University of Technology, No.2 Linggong Road, Dalian, Liaoning, 116024, P.R.China} \\
{\small $^{2}$ LAPTh, Universit$\rm \acute{e}$ de Savoie et CNRS, BP110, F-74941 Annecy-le-Vieux Cedex, France}\\
}
\date{}
\maketitle

\vspace{-0.7cm}
\begin{abstract}

In this paper, we study the anomalous flavor changing neutral current Yukawa interactions
between the top quark, the Higgs boson, and either an up or charm quark ($\rm tqH, q=u, c$).
We probe these couplings in $\rm e^- p\rightarrow \nu_e \bar{t} \rightarrow \nu_e H \bar{q}$
and the channel $\rm e^- p \rightarrow \nu_e H b$. Both channels are induced
by charged current interactions through $\rm e^- p$ collision at the Large Hadron Electron Collider(LHeC).
We study the signatures with the Higgs decay modes
$\rm H\rightarrow \gamma\gamma, b\bar{b}$ and $\tau^+\tau^-$.
Our results show that the flavor changing couplings $\rm \kappa_{tqH}$ can be probed down to a value of
0.0162 in $\rm e^- p \rightarrow \nu_e \bar{t} \rightarrow \nu_e H \bar{q}$ with $\rm H\rightarrow b \bar{b}$ at a 14 TeV LHeC
with a 150 GeV electron beam and 200 $\rm fb^{-1}$ luminosity.
This value of the coupling corresponds to the branching ratio $\rm Br(t\rightarrow qH)=1.34 \times 10^{-4}$.

\vspace{-1.2cm} \vspace{2.0cm} \noindent
\end{abstract}

\newpage
\section{Introduction}
The discovery of the Higgs boson
at the Large Hadron Collider(LHC)\cite{SMHiggs_ATLAS}\cite{SMHiggs_CMS}
is a major step towards understanding the electroweak symmetry breaking (EWSB)
mechanism and marks a new era in  particle physics.
In order to ultimately establish the nature of the Higgs,
a precise measurement of the Higgs couplings to fermions and gauge
bosons as well as the Higgs self-coupling is needed.
These precision measurements will be some of the most important tasks for experiments
at the LHC and the future colliders.
According to the analyses of the ATLAS and CMS collaborations,
the couplings of the Higgs boson
have been measured with an overall precision of about $15\%$,
which means that there still remains some room for the existence new physics.
Besides the Higgs boson, the measurement of the top quark properties is also important.
It is the heaviest known elementary particle
which makes it an excellent candidate for new physics searches.
To probe new physics through the Higgs boson and top quark, the top-Higgs Yukawa couplings
are of special interest since they are sensitive to new flavor dynamics
beyond the Standard Model(SM) not too far above the electroweak scale.
Furthermore, top quark, as the heaviest SM fermion, it owns the strongest Yukawa coupling.
Among the Higgs couplings to quarks, the most promising place to
reveal new physics at high energy colliders are processes involving top quarks.

The mass of the top quark is heavier than that of the observed Higgs boson,
which makes the top quark flavor changing neutral current(FCNC) processes
$\rm t\rightarrow qH(q=u,c)$ kinematically accessible.
In the SM, processes that are induced by FCNC in top quark production or decay
are extremely suppressed by the Glashow-Iliopoulos-Maiani(G.I.M.) mechanism\cite{GIM_mechanism}
according to SM computation, with decay rates of the order of $10^{-10}$ or below.
However, new physics scenarios, such as
the minimal supersymmetric model (MSSM) with/without R-parity
Violating\cite{MSSM_tqH_1}\cite{MSSM_tqH_2}\cite{MSSM_tqH_3}\cite{MSSM_tqH_4}\cite{MSSM_tqH_5}\cite{MSSM_tqH_6}\cite{MSSM_tqH_7}\cite{MSSM_tqH_8},
two-Higgs-Doublet Model (2HDM) \cite{2HDM_tqH_1}\cite{2HDM_tqH_2}\cite{2HDM_tqH_3}\cite{2HDM_tqH_4},
Warped Extra Dimensions\cite{WED_tqH_1}\cite{WED_tqH_2},
Alternative Left-Right symmetric Models (ALRM)\cite{ALRM_tqH},
Little Higgs with T parity (LHT)\cite{LHT_tqH_1}, etc,
could enhance the FCNC rates by several orders of magnitude,
thus making them detectable using current experimental data.
Therefore, studying the top-Higgs FCNC interactions is important both from a theoretical  as well as
an experimental perspective.

Up to now, the searches for $\rm t\rightarrow qH$ have been investigated
experimentally at the LHC which gives the strong limits on the top-Higgs FCNC couplings.
Among them, the most stringent constraint of $\rm Br(t\rightarrow cH)<0.56\%$,
$\rm Br(t\rightarrow uH)<0.45\%$ at $95\%$
confidence level (C.L.) was reported by the CMS collaboration from a combination of the multilepton
channel and the diphoton plus lepton channel\cite{FCNC_limit_tHq_CMS}.
While an upper limit is set on the $\rm t\rightarrow cH$ branching ratio of $0.79\%$ at the $95\%$ confidence level
by the ATLAS collaboration\cite{FCNC_limit_tHq_ATLAS}\cite{FCNC_limit_tHq_ATLAS1}.
Except for the widely studied $\rm t\rightarrow qH$ decays,
the importance of the single top Higgs associated production has been also emphasized
in the recent theoretical studies especially at the LHC
\cite{EFT_tqH_1}\cite{EFT_tqH_2}\cite{EFT_tqH_3}\cite{HL_LHC_limit}\cite{DipoleM_tqH}\cite{EFT_tqH_NLO_1}\cite{EFT_tqH_NLO_2}\cite{tqH_old}\cite{tqH_new}\cite{SM_tqH_multileptons}.
In our present paper, we study the anomalous FCNC Yukawa interactions
between the top quark, the Higgs boson, and either an up or charm quark ($\rm tqH, q=u,c$)
at the Large Hadron Electron Collider(LHeC).
The LHeC kinematic range exceeds HERA's by a factor of about 20, due to
the combination of a 7 TeV or higher proton beam from the LHC and a new 60 GeV to 150 GeV electron beam.
Its luminosity is projected to be as high as possibly $\rm 10^{34} cm^{-2} s^{-1}$,
with a default design value of $\rm 10^{33} cm^{-2} s^{-1}$.
This is almost a thousand times higher than HERA's luminosity, which gives
the LHeC the potential of a precision measurement Higgs production facility and enables a very large
variety of new measurements and searches to be conducted.
Typically we choose two channels to study the anomalous FCNC Yukawa interactions at the LHeC.
One is the channel $\rm e^- p\rightarrow \nu_e \bar{t} \rightarrow \nu_e H \bar{q}$
with $\rm q=u, c$ and the other is the channel $\rm e^- p \rightarrow \nu_e H b$. Both channels are
charged current (CC) interaction processes induced through $\rm e^- p$ collision at the LHeC.

Our paper is organized as follows: we build the calculation framework
in Section 2 including a brief introduction to the anomalous flavor changing $\rm tqH$ couplings
and our selected production channels.
Section 3 is arranged to present the numerical results as well as the signal and background analysis.
Typically, the $\rm H\rightarrow \gamma\gamma, b\bar{b}, \tau^+\tau^-$ decay modes are taken into account.
In Section 4 we present  bounds on anomalous $\rm tqH$ couplings at the future LHeC.
Finally we summarize our conclusions in the last section.

\section{Calculation Framework}

\subsection{Flavor Changing $\rm tqH$ Couplings}

Considering the FCNC Yukawa interactions, the SM Lagrangian can be extended simply by allowing the following terms,
\begin{eqnarray}\label{lagrangian}
\rm {\cal L} = \kappa_{tuH} \bar{t}uH + \kappa_{tcH} \bar{t}cH + h.c.,
\end{eqnarray}
where the real parameters $\rm \kappa_{tuH}$ and $\rm \kappa_{tcH}$ denote the flavor changing couplings
of Higgs to up-type quarks.  Now we have $\rm m_t-m_h$ larger than $\rm m_c, m_u, m_b$.
In addition to the usual top decay mode $\rm t\rightarrow W^\pm b$, the top quark can also decay into a charm or up quark
associated with a Higgs boson.  Therefore, the total decay width of the top-quark $\Gamma_t$ is
\begin{eqnarray}
\rm \Gamma_t = \Gamma^{SM}_{t\rightarrow W^-b}+\Gamma_{t\rightarrow cH}+\Gamma_{t\rightarrow uH}.
\end{eqnarray}
The decay width of the dominant top quark decay mode $\rm t\rightarrow W^-b$ at the LO and the NLO
could be found in ref\cite{twb_NLO}. It is given below
\begin{eqnarray}\label{GammaT} \nonumber
\rm \Gamma^{SM}_{t\rightarrow W^-b}=\Gamma_0(t\rightarrow W^-b) \{ 1+\frac{2\alpha_s}{3\pi}
 [2(\frac{(1-\beta^2_W)(2 \beta^2_W-1)(\beta^2_W-2)}{\beta^4_W(3-2\beta^2_W)}) \mbox{ln}(1-\beta^2_W)  \\
\rm - \frac{9-4\beta^2_W}{3-2\beta^2_W} \mbox{ln}\beta^2_W +2Li_2(\beta^2_W)
 -2Li_2(1-\beta^2_W) - \frac{6\beta^4_W-3 \beta^2_W -8}{2 \beta^2_W (3-2\beta^2_W)} -\pi^2 ] \}
\end{eqnarray}
where $\rm \Gamma_0(t\rightarrow W^-b)=\frac{G_F m_t^3}{8\sqrt{2}\pi}|V_{tb}|^2 \beta^4_W (3-2\beta^2_W)$
is the LO decay width and $\rm \beta_W=(1-m_W^2/m_t^2)^{\frac{1}{2}}$
is the velocity of the W boson in the top quark rest frame. $\rm G_F$ is the fermi constant.
The $\rm t\rightarrow u(c)H$ partial decay width is given as\cite{decay_tqH}
\begin{eqnarray}
\rm \Gamma_{t\rightarrow u(c)H}=\frac{\kappa^2_{tu(c)H}}{16\pi}  m_t  [(\tau_{u(c)}+1)^2-\tau^2_{H}]
\sqrt{1-(\tau_{H}-\tau_{u(c)})^2}\sqrt{1-(\tau_{H}+\tau_{u(c)})^2}
\end{eqnarray}
where $\rm \tau_H=\frac{m_H}{m_t}$, $\rm \tau_{u(c)}=\frac{m_{u(c)}}{m_t}$.
The leading order branching ration for $\rm t \rightarrow qH$ is then given by
\begin{eqnarray}
\rm Br(t\rightarrow u(c)H) = \frac{\kappa_{tu(c)H}^2}{\sqrt{2} G_F m_t^2 } \frac{(1-m_H^2/m_t^2)^2}{(1-m_W^2/m_t^2)^2(1+2m_W^2/m_t^2)}\approx 0.512\kappa_{tu(c)H}^2.
\end{eqnarray}
Here the Higgs boson and the top quark masses are chosen to be
$\rm m_H=125.7\ GeV$ and $\rm m_t=173.21\ GeV$ respectively.
Similar to the top quark decay, the new interactions 
affect also the width of the Higgs though the additional decay into an off-shell top
that subsequently leads to a single $W$ decay of the Higgs, namely $\rm H\rightarrow u(c) (t^*\rightarrow Wb)$
where $\rm t^*$ denotes off-shell top quark. Therefore we get
\begin{eqnarray}
\rm \Gamma_H=\Gamma_{H}^{SM} +
\Gamma_{H\rightarrow u(\bar{t}^*\rightarrow \bar{b}W^-)} +
\Gamma_{H\rightarrow \bar{u}(t^*\rightarrow bW^+)}+
\Gamma_{H\rightarrow c(\bar{t}^*\rightarrow \bar{b}W^-)} +
\Gamma_{H\rightarrow \bar{c}(t^*\rightarrow bW^+)}
 \end{eqnarray}
where $\rm \Gamma_{H}^{SM}$ is the normal Higgs decay width in SM.
While other terms related to the Higgs boson three-body decays are
numerically estimated following ref\cite{EFT_tqH_1}.

The stringent constraints on the anomalous
FCNC couplings are set exploiting  the experimental data of
 the CMS and ATLAS
Collaborations\cite{FCNC_limit_tHq_CMS}\cite{FCNC_limit_tHq_ATLAS}\cite{FCNC_limit_tHq_ATLAS1}.
Theoretically, many other phenomenological studies are performed based on these experimental data.
The analysis of ref\cite{EFT_tqH_2} emphasizes
the importance of anomalous single top plus Higgs production at the LHC deriving the $95\%$ C.L. upper limits
$\rm Br (t \rightarrow cH) < 0.15\%$ and $\rm Br (t\rightarrow uH) < 0.19\%$.
Ref\cite{EFT_tqH_3} studies the single top and Higgs associated production $\rm pp\rightarrow tHj$
in the presence of top-Higgs FCNC couplings at the LHC,
giving the upper limits
as $\rm Br(t\rightarrow qH)< 0.12\%$,
$\rm Br (t\rightarrow uH) < 0.26\%$ and $\rm Br (t \rightarrow cH) < 0.23\%$
with an integrated luminosity of 3000 $\rm fb^{-1}$ at $\rm \sqrt{s} = 14TeV$.
Ref\cite{HL_LHC_limit} quotes a $95\%$ C.L. limit Sensitivity in the
$\rm tt\rightarrow Wb+hq\rightarrow \ell \nu b + \ell\ell(\gamma\gamma) q$
final state of $\rm Br(t\rightarrow qH)<5(2)\times 10^{-4}$
with an integrated luminosity of 300(3000) $\rm fb^{-1}$ at $\rm \sqrt{s} = 14\ TeV$.
As can be seen the upper limits on the flavour changing top quark decays
can be significantly improved as expected at a High Lumi(HL)-LHC.
Ref\cite{DipoleM_tqH} derives model-independent constraints on the tcH and tuH couplings
that arise from the bounds on hadronic electric dipole moments.
Refs\cite{EFT_tqH_NLO_1} and \cite{EFT_tqH_NLO_2}
study the top quark decay into Higgs boson, a light quark and top Higgs associated production including the next-to-leading order QCD effects. Other related publications  can be found, for example, in refs\cite{tqH_old}\cite{tqH_new}\cite{SM_tqH_multileptons}\cite{SM_tqH_ChargeRatio}, etc.

\subsection{The Processes}

Now we turn to study the selected production processes where the effect of the flavor changing couplings could be significant.

\subsubsection{$\rm e^- p \rightarrow \nu_e \bar{t} \rightarrow \nu_e H \bar{q}$ Channel}

\begin{figure}[hbtp]
\vspace{-5cm}
\centering
\includegraphics[scale=0.6]{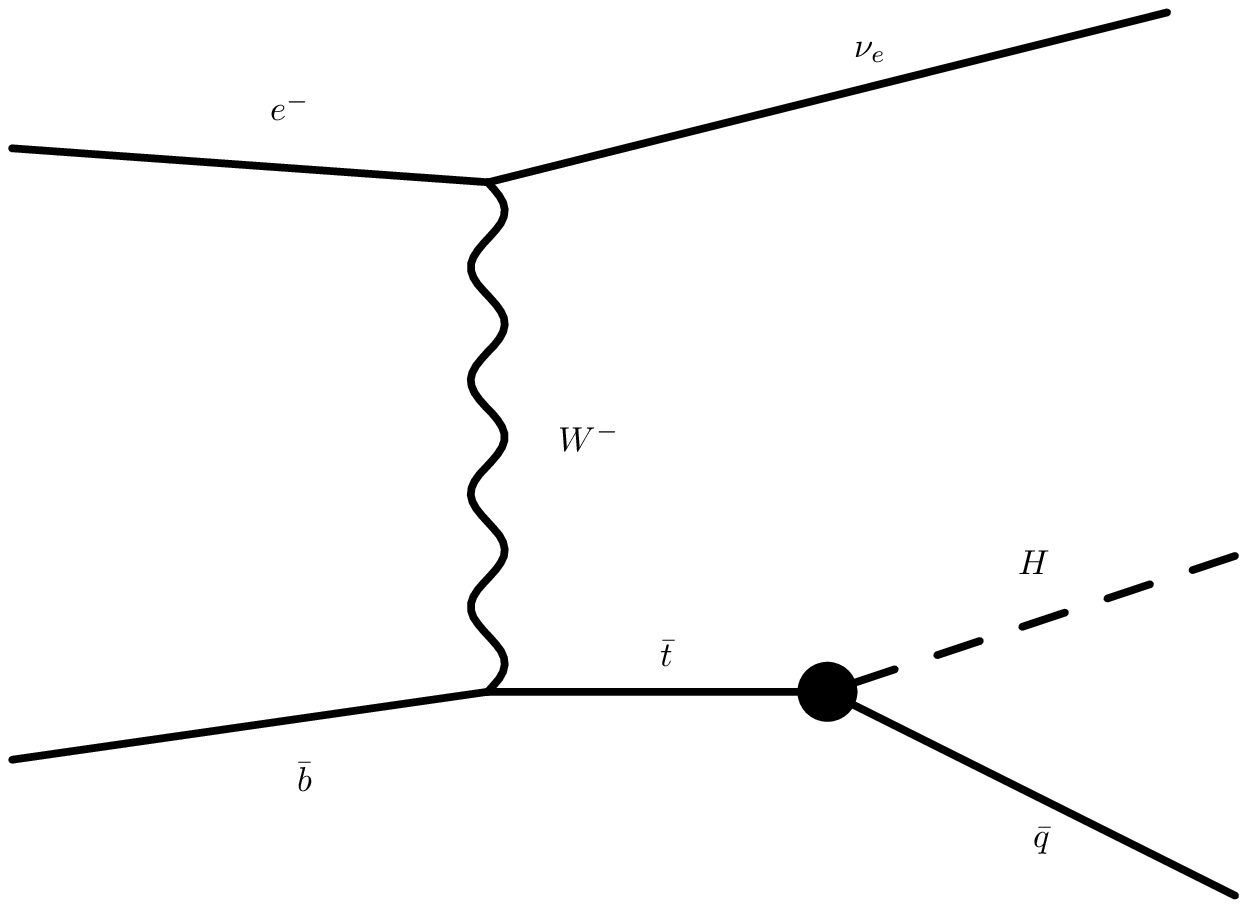}
\vspace{-7cm}
\caption{\label{feynman1} Partonic Feynman diagrams for
$\rm e^- p\rightarrow \nu_e \bar{t}\rightarrow \nu_e H \bar{q}$ with q=u ,c.
Black blobs represent the anomalous $\rm tqH$ couplings parameterized by Eq.(\ref{lagrangian}).}
\end{figure}

The first channel we will consider is $\rm e^- p\rightarrow \nu_e \bar{t}\rightarrow \nu_e H \bar{q}$ production.
The parton level signal process at the LHeC can be expressed as
\begin{eqnarray}
\rm e^-(p_1) + \bar{b}(p_2) \rightarrow  \nu_e + \bar{t} \rightarrow  \nu_e(p_3) + H(p_4) + \bar{q}(p_5)
\end{eqnarray}
with $\rm q=u, c$ and $\rm p_i$ are the four-momentum of initial and final particles, respectively.
The Feynman diagram for the partonic process is depicted in Fig.\ref{feynman1}.
The flavor changing vertex proportional to the flavor changing coupling
$\rm \kappa_{tqH}$ occurs via the single top production with its following decay to Higgs plus u or c quark,
where this single top quark is induced by the collision of b quark from the proton
with the $\rm W^-$ boson emission from the electron beam.
We thus expect the cross sections for these processes to be proportional to
$\rm c\kappa_{tqH}^2$ where c is some related constants.
The parent level signal process $\rm e^- p \rightarrow \nu_e H \bar{q} +X$, the kinematic distributions
and integrated cross sections can then be obtained by convoluting the parton level process with
the parton distribution function (PDF)\cite{LHeCPDF} of quark in the proton,
\begin{eqnarray}
\rm d\sigma (e^- p \rightarrow \nu_e H \bar{q} +X) = \int dx \ G_{\bar{b}/P}\ (x,\mu_f) d\hat\sigma
(e^- \bar{b} \rightarrow \nu_e \bar{t} \rightarrow  \nu_e H \bar{q}, \sqrt{\hat{s}}),
\end{eqnarray}
where $\rm \sqrt{\hat{s}}=2\sqrt{xE_e E_p}$ is the center-of-mass(c.m.) colliding energy
and x is the momentum fraction of anti-b quark from proton.

\subsubsection{$\rm e^- p \rightarrow \nu_e H b$ Channel}

\begin{figure}[hbtp]
\vspace{-5cm}
\centering
\includegraphics[scale=0.6]{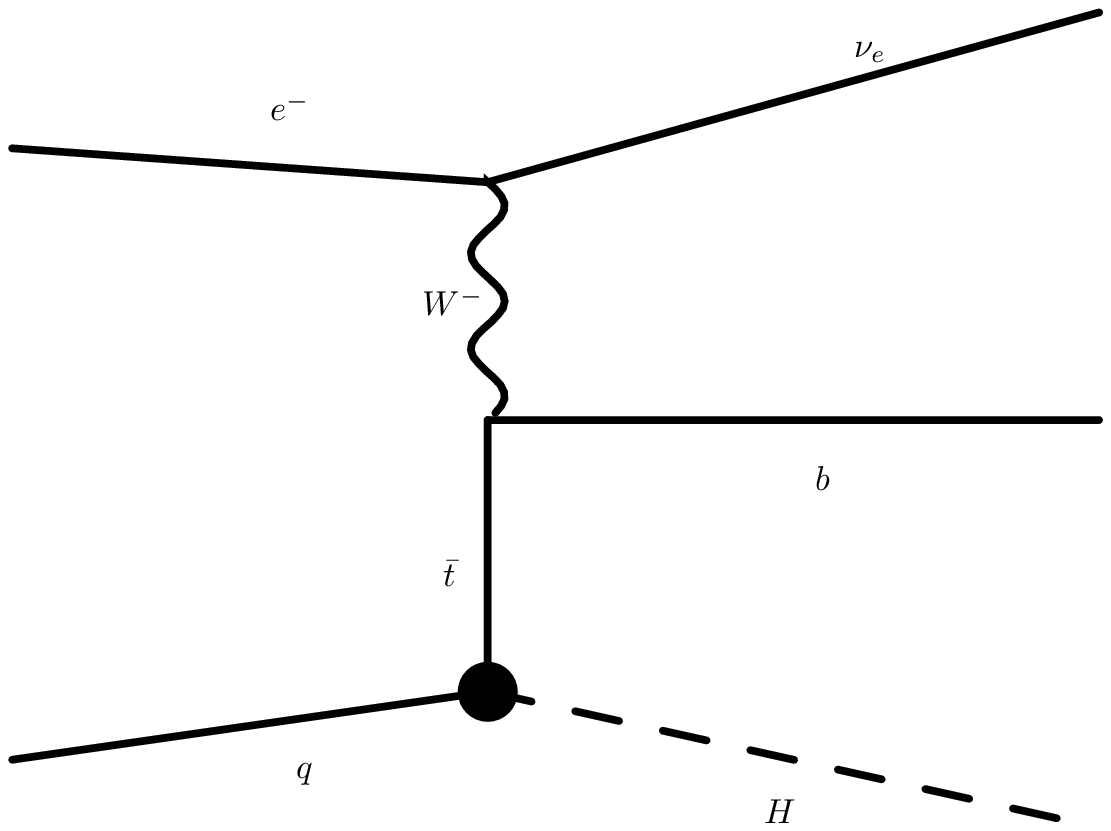}
\vspace{-7cm}
\caption{\label{feynman2} Partonic Feynman diagrams for
$\rm e^- p \rightarrow \nu_e H b$.
Black blobs represent the anomalous $\rm tqH$ couplings parameterized by Eq.(\ref{lagrangian}).}
\end{figure}

The second channel we considered is $\rm e^- p \rightarrow  \nu_e H b$ production.
The parton level signal process at the LHeC can be expressed as
\begin{eqnarray}
\rm e^-(p_1) + q(p_2) \rightarrow  \nu_e(p_3) + H(p_4) + b(p_5).
\end{eqnarray}
The Feynman diagram for the partonic process is depicted in Fig.\ref{feynman2}.
The FCNC  top-Higgs  Yukawa couplings are deduced from the initial state u(c)-quarks from the proton
collision with the anti-top quark from the $\rm Wtb$ coupling.
Similarly, the parent level signal process $\rm e^- p \rightarrow \nu_e H b +X$ is present as
\begin{eqnarray}
\rm d\sigma (e^- p \rightarrow \nu_e H b +X) = \int dx \ G_{q/P}\ (x,\mu_f) d\hat\sigma
(e^- q \rightarrow   \nu_e H b, \sqrt{\hat{s}})
\end{eqnarray}
where $\rm q=u, c$ and $\sqrt{\hat{s}}$ is again the c.m. colliding energy at the LHeC.

\subsubsection{Charged Current and Neutral Current Production at the LHeC}

The two channels $\rm e^- p\rightarrow \nu_e \bar{t}\rightarrow \nu_e H \bar{q}$
and $\rm e^- p \rightarrow  \nu_e H b$ that we have presented are  charged current (CC) processes
where the CC production leads to a top-beauty associated production through $\rm W^-$ boson emission from
the initial electron.  In addition to CC production,  the flavor changing Yukawa couplings can also be produced
through neutral current (NC) productions. In NC it gives rise to pair production of top-antitop quarks from
a neutral photon/Z boson emission from the initial electron.
A comparison of the cross sections of these CC and NC production channels  including the anomalous
FCNC top-Higgs Yukawa couplings is presented in Tab.\ref{ccnc}.
Here the input parameters and the very basic kinematical cuts will be presented in our following discussion.

\begin{table}[htb]
\begin{center}
\begin{tabular}{c c }
\hline\hline
 \multicolumn{2}{c}{($\rm \sqrt{s}_{e^-}$,$\rm \sqrt{s}_{p}$)= (150\ [GeV], 14\ [TeV])  }   \\
 channels & $\rm \sigma(\kappa_{tuH}=0.1)$[fb] \\
 \hline
  $\rm e^- p\rightarrow \nu_e \bar{t} \rightarrow \nu_e H \bar{q}$ &  41.64\\
  $\rm e^- p\rightarrow  \nu_e H b$ &1.987 \\
  $\rm e^- p\rightarrow e^- Ht$  & 0.616\\
  $\rm e^- p \rightarrow e^- H q W$ & 0.901\\
 \hline\hline
 \end{tabular}
 \end{center}
 \caption{\label{ccnc}
 A comparison of the cross sections of CC and NC production channels  including the anomalous
FCNC top-Higgs Yukawa couplings with $\rm \kappa_{tuH}=0.1$.}
\end{table}

 \begin{figure}[htb]
 \center{
 \includegraphics[scale=0.8]  {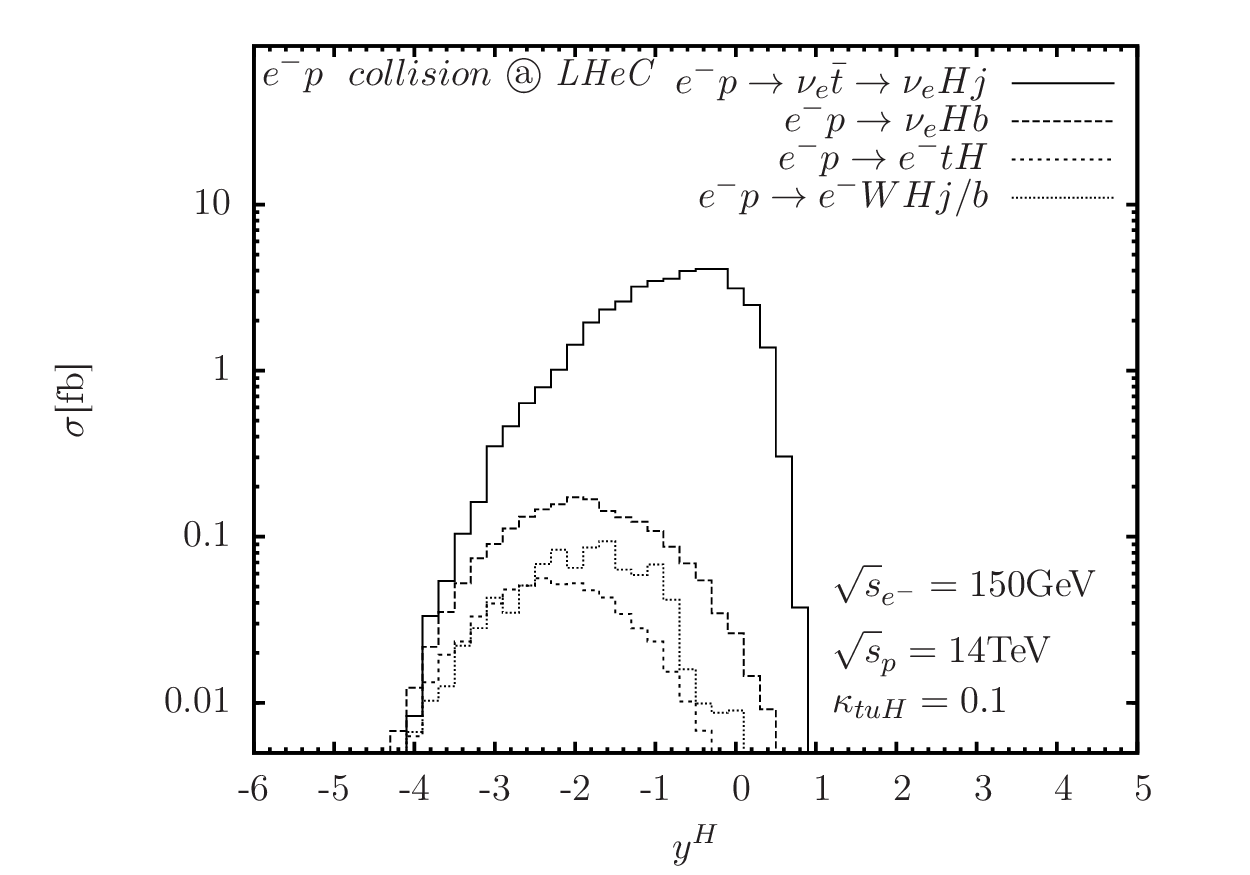}
 }\caption{\label{yh}
 The rapidity distributions of the Higgs boson through different channels
 including $\rm e^- p\rightarrow \nu_e \bar{t} \rightarrow \nu_e H \bar{q}$,
$\rm e^- p\rightarrow \nu_e H b$, $\rm e^- p\rightarrow e^- Ht$ and
$\rm e^- p \rightarrow e^- H q W$ productions.}
 \end{figure}

From Tab.\ref{ccnc}, we see that the largest production is CC
$\rm e^- p\rightarrow \nu_e \bar{t} \rightarrow \nu_e H \bar{q}$ production. For $\rm \kappa_{tuH}=0.1$ it
is more than 10 times larger than the sum of the other channels.
Different from the CC production which leads to a top-beauty final state,
the NC production gives rise to pair produced top-antitop quarks.
The NC productions are small, but still sizeable
at the LHeC especially when the polarized electron beam is considered.
Furthermore, in sharp contrast to the LHC, the absence of pile-up
and underlying event effects at the LHeC,
high rates of single anti-top production is expected to
providing a better insight through these production channels.
The rapidity distributions of the Higgs boson
through different channels are given in Fig.\ref{yh}.
In our paper, we only consider the CC interactions
which dominate over all the other production mechanisms.
This includes $\rm e^- p\rightarrow \nu_e \bar{t} \rightarrow \nu_e H \bar{q}$
and $\rm e^- p\rightarrow \nu_e H b$ channels.
Looking at Tab.\ref{ccnc}, we also find that the cross section of the former channel is larger
than that of the latter one by roughly a factor of 20.
At first sight this seems odd, because the transition
$\rm e^- p(\bar{b}) \rightarrow \nu_e \bar{t} \rightarrow \nu_e H \bar{q}$
involves an (anti)bottom-quark PDF, while the transition of $\rm e^- p(q)\rightarrow \nu_e H b$ does not.
One  is therefore tempted to think that the cross section of
$\rm e^- p\rightarrow \nu_e \bar{t} \rightarrow \nu_e H \bar{q} $ is smaller than
that of $\rm e^- p\rightarrow \nu_e H b $.
However, this naive assertion is incorrect, because for
$\rm e^- p\rightarrow \nu_e \bar{t} \rightarrow \nu_e H \bar{q}$
the internal (anti)top is exchanged in the s-channel, while in the case
$\rm e^- p\rightarrow \nu_e H b$ the top appears in a t-channel exchange.
The PDF suppression is thus over compensated by an on-shell enhancement.

\section{Results}

\subsection{Input Parameters}

We take the input parameters as \cite{2014PDG}
$\rm \alpha_{ew}(m^2_Z)^{-1}|_{\overline{MS}}=127.9$,
$\rm G_F=1.166370\times 10^{-5}\ GeV^{-2}$,
$\rm m_Z=91.1876\ GeV$,
so we have $\rm m_W=79.82436\ GeV$ and
$\rm \sin^2\theta_W=1-(m_W/m_Z)^2=0.233699$.
For the strong coupling constant we take $\rm \alpha_s = 0.1184 $.
Throughout this paper, we set the quark masses as $\rm m_u=m_d=m_c=m_s=0\ GeV$ and
$\rm m_b=4.18\ GeV$. The top quark mass is set to $\rm m_t=173.21\ GeV$
with its width  $\rm \Gamma_t=1.3604\ GeV$ when $\rm \kappa_{tuH}=0.1$.
For the leptons, we keep $\rm m_e=m_{\mu}=0\ GeV$, and $\rm m_\tau=1.77682\ GeV$.
We do not consider the contribution from small CKM matrix $\rm V_{qq'}$
where q and $\rm q'$ are not the same generation.
For the mass of the Higgs boson, we take $\rm m_H=125.7\ GeV$
with the SM width to be $\rm \Gamma^{SM}_H=4.3\ MeV$.
The partonic cross sections are convoluted with CTEQ6L1\cite{CTEQ6} parton distribution
functions (PDF)  keeping factorization and renormalization scale $\rm \mu_f=\mu_r = m_t$.
For the LHeC colliding energy, we consider the future 14 TeV proton at future LHC
and an energetic new electron beam with the energies of 150 GeV\cite{LHeC_Collider1}\cite{LHeC_Collider2}.
The luminosity is taken to be a running parameter.
The FCNC couplings are chosen to be $\rm \kappa_{tuH}=0.1$ and $\rm \kappa_{tcH}=0$ for simplicity.
This set of parameters will be used as default unless being stated otherwise.

\subsection{Kinematic Cuts}

The event reconstruction is still based on a parameterised, generic
LHC-style detector.  The general acceptance cuts  in the lab frame for the events are:
\begin{eqnarray} \label{basiccuts}\nonumber
&&\rm p_T^{jet} \geq 25~{\rm GeV}, p_T^b \geq 25~{\rm GeV},p_T^\gamma \geq 25~{\rm GeV},
p_T^{\ell} \geq 25~{\rm GeV}, \slashed{E}_T^{miss} \geq 25~{\rm GeV}, \\\nonumber
&&\rm |\eta^{jet}|<5,  |\eta^b|<2.5, |\eta^\gamma|<2.5, |\eta^\ell|<2.5, \\\nonumber
&&\rm \Delta R (j j)>0.4,  \Delta R (bb)>0.4, \Delta R (\ell\ell)>0.4, \Delta R (\gamma\gamma)>0.4,  \Delta R(\gamma \ell)>0.4 \\
&&\rm \Delta R (j b)>0.4, \Delta R (\ell j)>0.4 , \Delta R (\ell b)>0.4 , \Delta R (\gamma b)>0.4, \Delta R(\gamma j)>0.4
\end{eqnarray}
where $\rm \Delta R = \sqrt{\Delta \Phi^2 + \Delta \eta^2}$ is the separation
in the rapidity-azimuth plane, $\rm p_T^{jet, \ell, \gamma}$ are the transverse momentum
of jets (refer as j), leptons and photons while $\rm  \slashed E_T^{miss}$ is the missing transverse momentum.
We stress here that cuts in Eq.({\ref{basiccuts}}) are the very basic ones and might be changed
later in our following discussion.

\subsection{Decay Modes and Backgrounds}

\subsubsection{$\rm e^- p \rightarrow \nu_e \bar{t} \rightarrow \nu_e H \bar{q}$
Channel with $\rm H\rightarrow \gamma\gamma(b\bar{b},\tau^+\tau^-) $ Decay Modes}

Lets first consider the $\rm e^- p \rightarrow \nu_e \bar{t} \rightarrow \nu_e H \bar{q}$
Channel with $\rm H\rightarrow \gamma\gamma$ decay mode.
The considered signal production can be written as
\begin{eqnarray}
\rm e^- p \rightarrow \nu_e \bar{t} \rightarrow \nu_e H \bar{q} \rightarrow  \nu_e \gamma\gamma \bar{q}
\end{eqnarray}
with $\rm q=u, c$. Since in our calculation we take the anomalous FCNC couplings to be
$\rm \kappa_{tuH}=0.1$ and $\rm \kappa_{tcH}=0$ for simplicity, only $\rm q=u$ contributes.
Higgs decays to pairs of photons are simulated using MadGraph\cite{MadGraph} where the implementation of
the effective $\rm H\gamma\gamma$ interaction is adopted
\cite{EFT_Hvv}.
For simplicity one can also multiply the production cross sections with the Higgs branching ratio
corresponding to the final state.
As can be seen, in this case, the studied topology of our signal  gives rise to the
$\rm jet+\slashed E_T+ diphoton$ signature characterized by one jet,
a missing transverse momentum ($\rm \slashed E_T$) from the undetected neutrino and a diphoton signal
appearing as a narrow resonance centered around the Higgs boson mass.
The irreducible background comes from the SM process
$\rm e^- p \rightarrow  \nu_e \gamma\gamma \bar{q}$ which yields the identical final states with the signal.
These backgrounds mainly come from the production of W boson with double photon production
through  $\rm WW\gamma\gamma$, $\rm W\gamma\gamma$ couplings or through
$\rm WW\rightarrow H\rightarrow\gamma\gamma$ decay  associated with jet emission.
The others come from jet production associated with emission of photons.
In order to obtain the anomalous FCNC tqH coupling effects,
we need to simulate all the signal contributions precisely
together with  these irreducible backgrounds as well as their interference.
The total cross section for these reactions thus can be split into three contributions
\begin{eqnarray}
\rm  \sigma=a_0+ a_1 \kappa_{tuH}  + a_2 \kappa_{tuH}^2
\end{eqnarray}
where $\rm a_0$ is the SM prediction, the term $\rm a_1$ linear in $\rm \kappa_{tuH}$
arises from the interference between SM and the anomalous amplitudes, whereas
the quadratic term $\rm a_2$ is the self-interference of the anomalous amplitudes.
Potentially reducible backgrounds come from various other SM processes
that yield different final states which are attributed to the $\rm jet+\slashed E_T+ diphoton$ signature
due to a misidentification of one or more of the final state objects.
For example, two light jets production with both jets faking a diphoton pair,
one jet one photon associated production with one jet faking a photon or leptons faking photons, etc.
The background arising from $\rm e^- p \rightarrow  \nu_e \bar{\nu_e} e^- \gamma  j$
is smaller than 1 percent of signal after applying all cuts and taking rejection factors into account.
We consider all these contributions and
take the jet faking a photon rate to be 0.001, the electron faking photon rate to be 0.062\cite{erfakerate} during data analysis.
Although the $\gamma\gamma$ decay channel has a small branching ratio, it has the advantage of
good resolution on the $\gamma\gamma$ resonance and is also free from the large QCD backgrounds.
Typically, we use a narrow invariant mass window $\rm |m_{\gamma\gamma} - m_H | < 5$ GeV
to further reduce the non-resonant backgrounds as well as
the jet such that the invariant mass of $\rm j\gamma\gamma$ system to be near mass of the top quark,
say, $\rm m_{j\gamma\gamma}$ belongs to the range $\rm [m_t-10, m_t+10]$ GeV.

\begin{figure}[hbtp]
\centering
\includegraphics[scale=0.6]{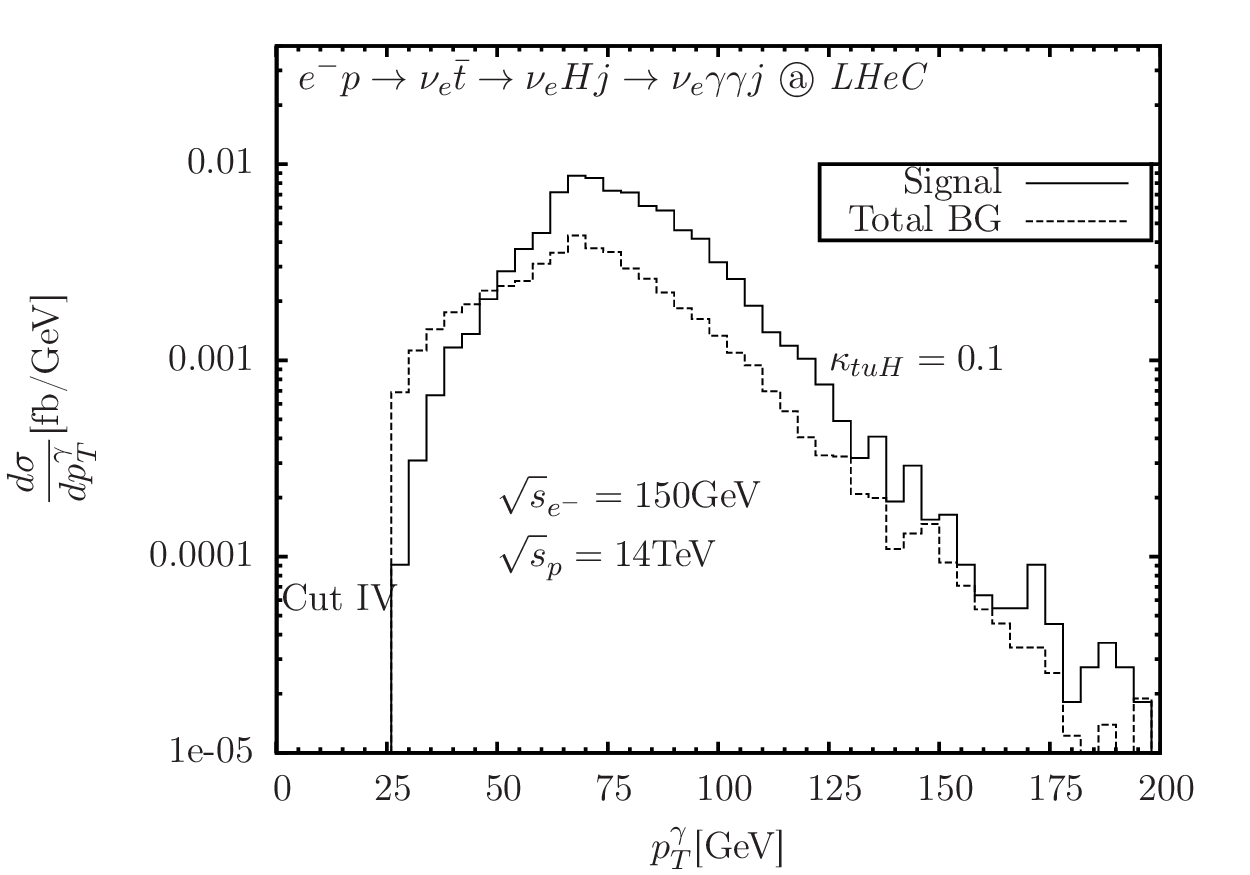}
\includegraphics[scale=0.6]{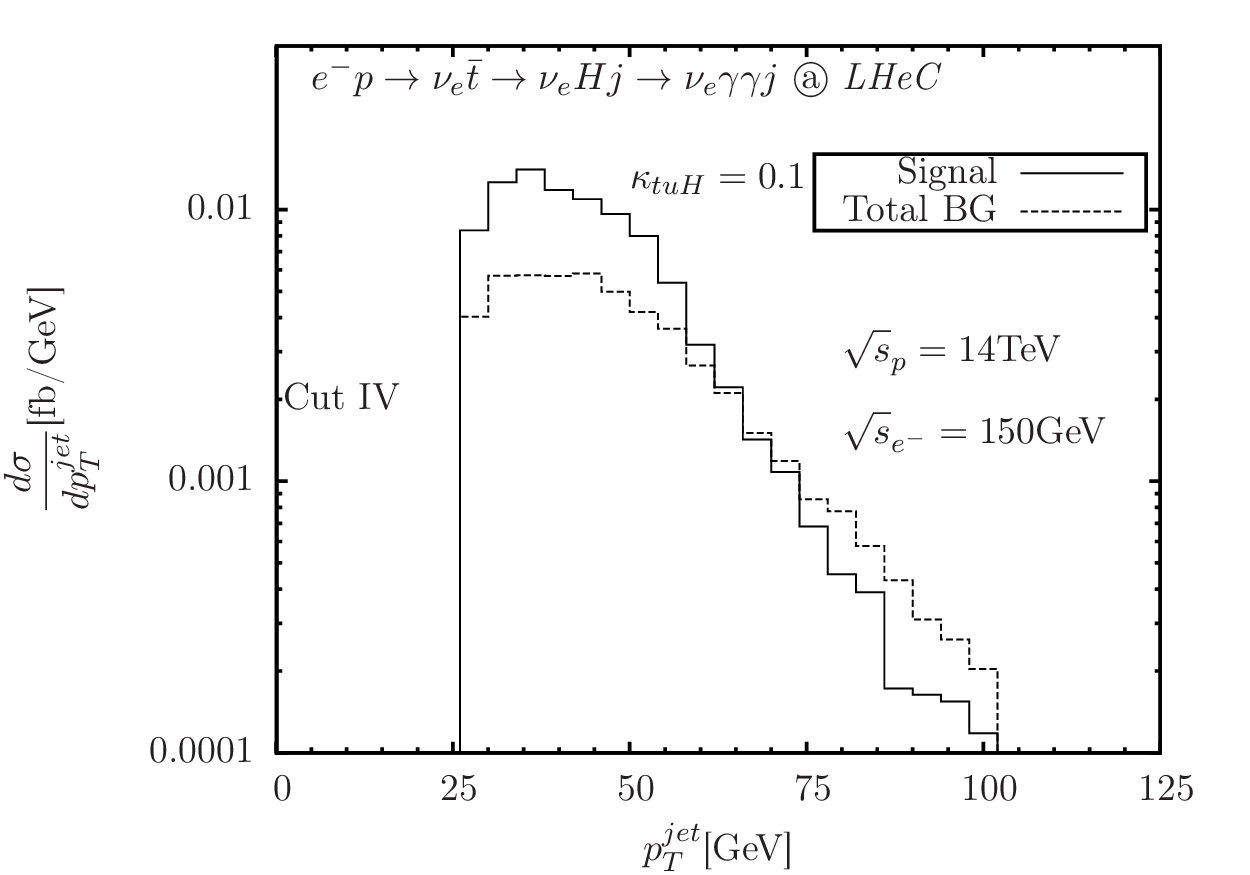}
\includegraphics[scale=0.6]{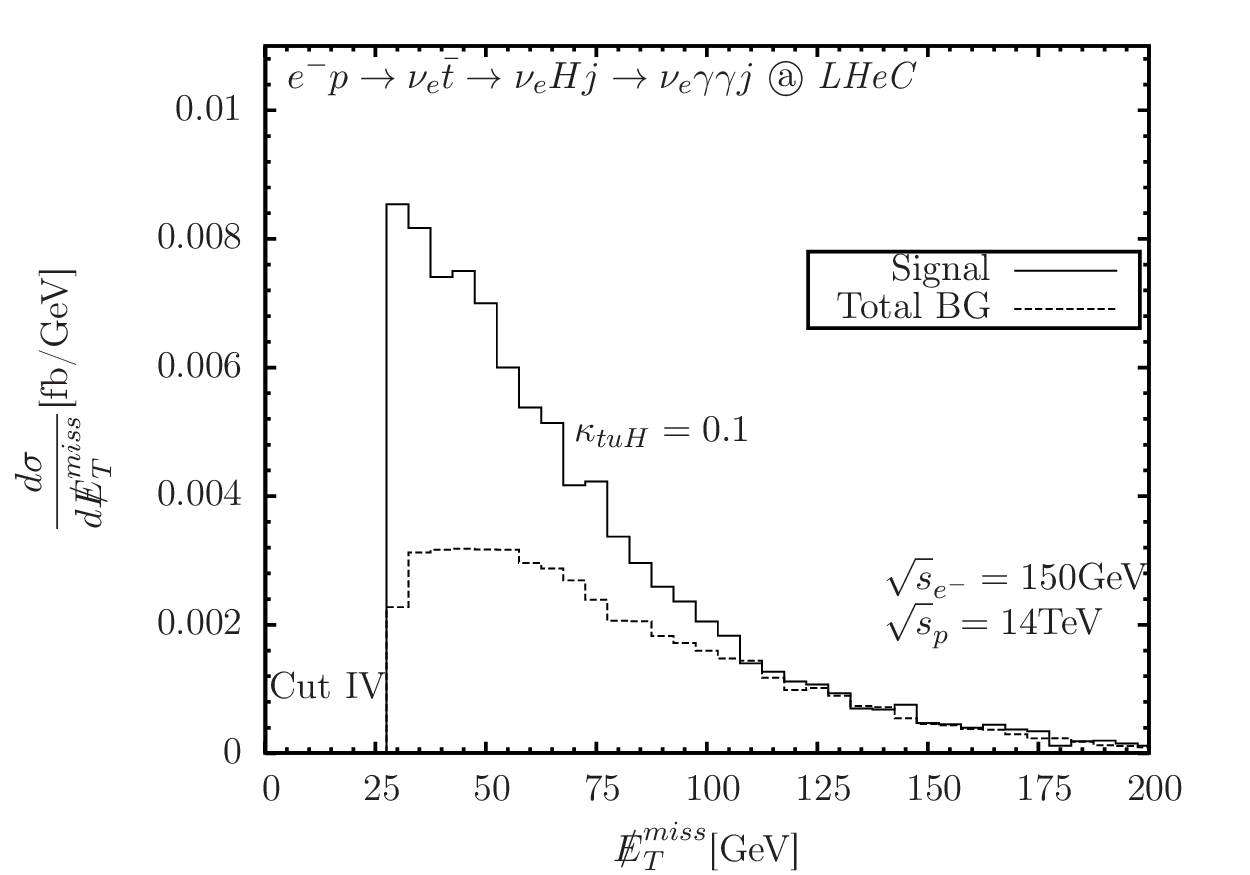}
\includegraphics[scale=0.6]{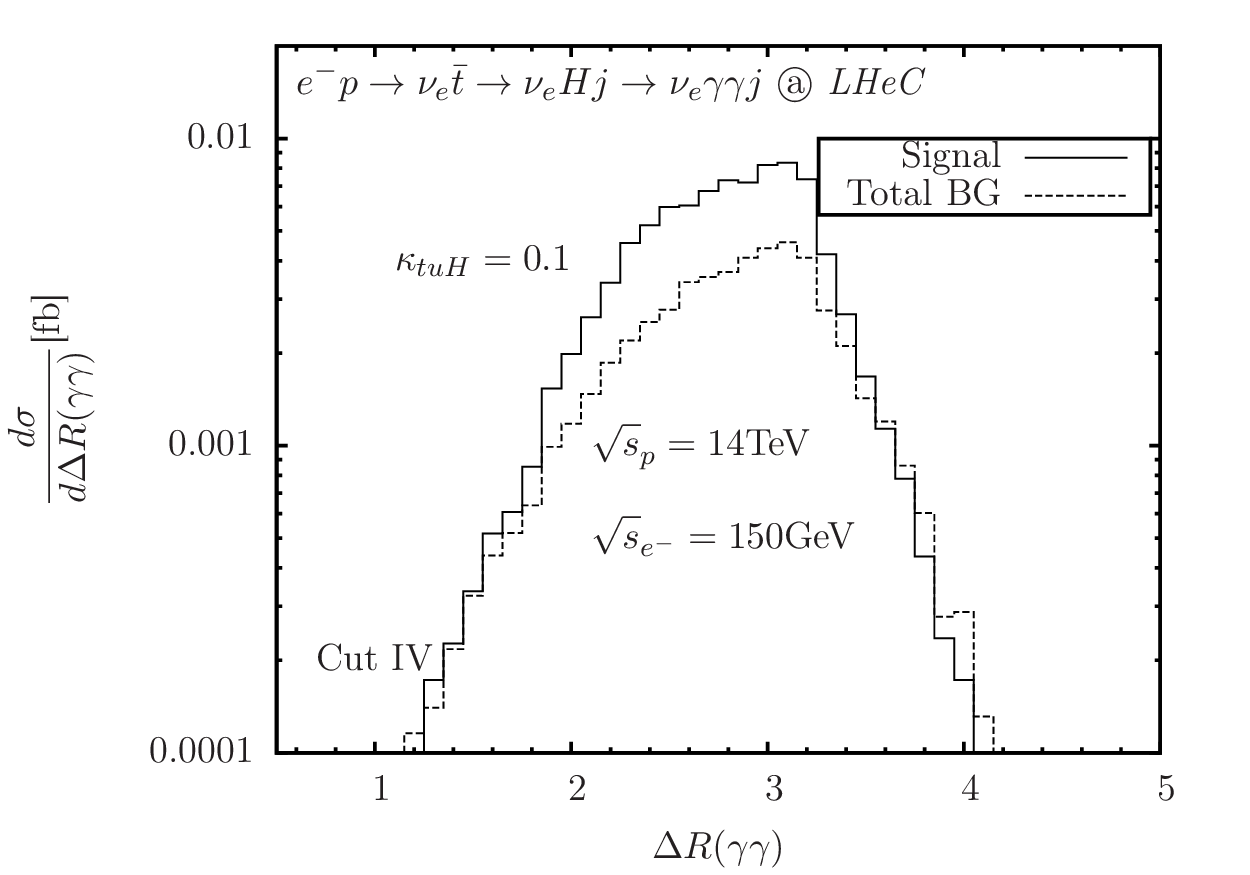}
\caption{\label{rrjdis}
The signal and total background transverse missing energy($\slashed E_T^{miss}$) distributions, transverse momentum ($\rm p_T^{\gamma,jet}$)
 distributions and $\Delta R (\gamma\gamma)$ distributions for $\rm e^- p \rightarrow  \nu_e \gamma \gamma j$
after considering Cut I-IV. The anomalous coupling is chosen to be $\rm \kappa_{tuH}=0.1$.
The rejection factors are taken into account.}
\end{figure}

\begin{table}[htb]
\begin{center}
\begin{tabular}{c c c c c c}
\hline\hline
\multirow{2}{2cm}{Decay} &\multirow{2}{1cm}{ [pb]}
& \multicolumn{4}{c}{ Cross sections for $\rm e^- p \rightarrow \nu_e \bar{t} \rightarrow \nu_e H \bar{q}$ channel($\rm \kappa_{tuH}=0.1$)} \\
&& Cut I & Cut II& Cut III  &  Cut IV  \\
 \hline
 \multirow{2}{2cm}{$\rm H\rightarrow \gamma\gamma $} & $\rm \sigma_{S}$  & 9.31$\times 10^{-5}$  &9.26$\times 10^{-5}$& 9.21$\times 10^{-5}$ &9.04$\times 10^{-5}$ \\
&  $\rm \sigma_{B}$ & $\rm 2.75\times 10^{-2}$& $\rm 1.35\times 10^{-2}$  & $\rm 1.02\times 10^{-3}$ &$\rm 5.29\times 10^{-5}$\\
& $\rm S/B$ & $\rm 3.39\times 10^{-3}$ & $\rm 6.86\times 10^{-3}$ & $\rm 9.03\times 10^{-2}$ & 1.71 \\
 \hline
 \multirow{2}{2cm}{$\rm H\rightarrow b\bar{b}$} &$\rm \sigma_{S}$ & $\rm 1.33\times 10^{-2}$ &$\rm 1.33\times 10^{-2}$ & $\rm 1.33\times 10^{-2}$ & $\rm 1.30\times 10^{-2}$ \\
&$\rm \sigma_{B}$ & $\rm 2.65\times 10^{-1}$ & $\rm 1.97\times 10^{-1}$  & $\rm 6.12\times 10^{-2}$ & $\rm 3.02\times 10^{-3}$ \\
&$\rm S/B$ & $\rm 5.02\times 10^{-2}$ & $\rm 6.75\times 10^{-2}$ & $\rm 2.17\times 10^{-1}$ & $\rm 4.30$ \\
 \hline
 \multirow{2}{2cm}{$\rm H\rightarrow \tau^+\tau^-$} &$\rm \sigma_{S}$ &$\rm 2.24\times 10^{-3}$&$\rm 2.23\times 10^{-3}$& $\rm 2.23\times 10^{-3}$ &$\rm 2.23\times 10^{-3}$ \\
&$\rm \sigma_{B}$ & $\rm 4.93\times 10^{-2}$  &$\rm 1.87\times 10^{-2}$ & $\rm 6.89\times 10^{-3}$ &$\rm 4.20\times 10^{-4}$\\
&$\rm S/B$ & $\rm 4.54\times 10^{-2}$ & $\rm 1.19\times 10^{-1}$ & $\rm 3.24\times 10^{-1}$ & $\rm 5.31$ \\
 \hline\hline
\end{tabular}
\end{center}
\caption{\label{SBXsectionrrj}
Signal and total Background cross sections for $\rm e^- p \rightarrow \nu_e \bar{t} \rightarrow \nu_e H \bar{q}$ channel
with different decay modes after the application of Cut I-IV. The rejection factors and b-tagging effects are taken into account in this table.}
\end{table}

We define some sets of kinematical cuts as bellow:
\begin{itemize}
 \item Cut I means the basic cuts present in Eq.(\ref{basiccuts});
 \item Cut II means the basic cuts plus $\rm 25<\slashed E_T^{miss}<300 GeV$, $\rm 25<p_T^{jet}<100 GeV$,
$\rm 25<p_T^\gamma<200 GeV$;
 \item Cut III means Cut II plus requiring the invariant mass of the diphoton pair  to be in the range [$\rm m_H-5$, $\rm m_H+5$ ] GeV;
 \item Cut IV means Cut III plus requiring the invariant mass of the diphoton and light jet system  to lie in the range [$\rm m_t-10$, $\rm m_t+10$] GeV.
\end{itemize}
In Tab.\ref{SBXsectionrrj}, we display the signal and the main background cross sections
for $\rm e^- p \rightarrow  \nu_e \gamma \gamma j$
after the application of Cut I-IV. The rejection factors and the b-tagging effects are already taken into account in this table,
where $\rm \sigma_{S}$ means the cross section for signal, $\rm \sigma_{B}$ for the background.
In Fig.\ref{rrjdis} we display the signal's and the total background's
transverse missing energy($\slashed E_T^{miss}$) distributions, transverse momentum ($\rm p_T^{\gamma,jet}$)
distributions and $\Delta R (\gamma\gamma)$ distributions for $\rm e^- p \rightarrow  \nu_e \gamma \gamma j$
in parton level after considering Cut I-IV. The anomalous coupling is chosen to be $\rm \kappa_{tqH}=0.1$.
The rejection factors are taken into account. We see that the anomalous FCNC tqH couplings
can enhance the SM production  to a level where it can be  detectable at future LHeC.
By a simple fit we get the final cross section to be
$\rm \sigma_{total} = 5.10\times 10^{-5} +5.21\times 10^{-5}\kappa_{tuH} + 8.63\times 10^{-3}\kappa^2_{tuH}$ [pb].

Now we consider the $\rm e^- p \rightarrow \nu_e \bar{t} \rightarrow \nu_e H \bar{q}$
channel with $\rm H\rightarrow b \bar{b}$ decay mode.
In this case, the signal production channel is characterized
by a missing energy from the undetected neutrino and a $\rm b \bar{b}$ pair associated with a light jet signal.
Still, the $\rm b \bar{b}$ pair signal is appearing as a narrow resonance centered around the Higgs boson mass.
The main background processes are $\rm e^- p \rightarrow \nu_e b \bar{b} j, bjj, \bar{b}jj$,
etc, with light jets faking b jets.
In our analysis, we assume a b-jet tagging efficiency of $\rm \epsilon_b=60\%$
and a corresponding mistagging rate of $\rm \epsilon_{light}=1\%$
for light jets (u, d, s quark or gluon) and $\rm \epsilon_c=10\%$ for a c-jet,
consistent with typical values assumed by the LHC experiments\cite{misjets}.
For this decay mode, we take Cut I, III, IV the same while Cut II  to be
the basic cuts plus $\rm 25<\slashed E_T^{miss}<400 GeV$, $\rm 25<p_T^b<200 GeV$,
$\rm 25<p_T^{jet}<140 GeV$ and  $\rm \Delta R(bj)<4$. For the background production, we also need that
the special cut for $\rm \nu_e \bar{b} jj$, $\rm \nu_e \bar{b} \bar{c} j$, etc,
with the light jets system not belongs to the range [$\rm m_W-10$, $\rm m_W+10$] GeV.
This cut will not affect the signal much but it will reduce the background obviously.
Finally we get the signal and total background to be 13 fb and 3.02 fb, respectively,
and we get the signal background ratio to be 4.3. The final cross section can be written as
$\rm \sigma_{total}=2.87\times 10^{-3} + 7.68\times 10^{-3}\kappa_{tuH} + 1.24\kappa^2_{tuH}$ [pb].

Finally we consider the $\rm \tau^+\tau^-$ decay mode in this production channel.
Our results show that $\rm e^- p \rightarrow \nu_e \bar{t} \rightarrow \nu_e H \bar{q}$
channel with $\rm H\rightarrow \tau^+ \tau^-$ decay mode
can be another good choice. With the four lists of cuts,
we take Cut I, III, IV the same while Cut II  to be
the basic cuts plus $\rm 25<\slashed E_T^{miss}<300 GeV$,
$\rm 25<p_T^j<100 GeV$, $\rm 25<p_T^\ell<200 GeV$
and $\rm \Delta R(\ell j)<4$. The total cross section can be parametrised as
$\rm \sigma_{total} = 3.96\times 10^{-4} + 1.30\times 10^{-3}\kappa_{tuH} + 2.13\times 10^{-1}\kappa^2_{tuH}$ [pb].
The cross sections of the above decay modes are presented
in Tab.\ref{SBXsectionrrj} with different sets of cuts.
We see that the $\rm H\rightarrow \gamma\gamma$ decay mode provides the smallest signal since
the branching ratio of $\rm H\rightarrow \gamma\gamma$ is quite small.
By applying the cuts, the background can be reduced to the same level.
For the $\rm H\rightarrow b\bar{b}, \tau^+\tau^-$ decay modes,
the signal can be 5 times larger than the
backgrounds, thus making the signal over background around 5 for $\rm \kappa_{tuH}$ equal 0.1.
The distributions of the signals and backgrounds are similar to Fig.\ref{rrjdis}
and we therefore do not display them.

\begin{table}[htb]
\begin{center}
\begin{tabular}{c c c c c c}
\hline\hline
\multirow{2}{2cm}{Decay} &\multirow{2}{1cm}{ [pb]}& \multicolumn{4}{c}{ Cross sections for $\rm e^- p\rightarrow \nu_e H b$ channel($\rm \kappa_{tuH}=0.1$)} \\
 &  & Cut I & Cut II& Cut III  &  Cut IV  \\
 \hline
 \multirow{3}{2cm}{$\rm H\rightarrow \gamma\gamma $} & $\rm \sigma_{S}$ & 1.66$\times 10^{-6}$  &1.24$\times 10^{-6}$& 1.23$\times 10^{-6}$ &0.80$\times 10^{-7}$ \\
 &  $\rm \sigma_{B}$ & 2.70$\times 10^{-4}$& 5.31$\times 10^{-5}$  & 1.99$\times 10^{-6}$ &3.32$\times 10^{-9}$\\
 & $\rm S/B$ &
$\rm 6.15\times 10^{-3}$& $\rm 2.34\times 10^{-2}$  & $\rm 6.18\times 10^{-1}$&$\rm 24.1$\\
 \hline
 \multirow{3}{2cm}{$\rm H\rightarrow b\bar{b}$} &$\rm \sigma_{S}$ & $\rm2.38\times 10^{-4}$  &$\rm2.36\times 10^{-4}$& $\rm2.16\times 10^{-4}$ &$\rm3.44\times 10^{-6}$ \\
  &$\rm \sigma_{B}$  &$\rm 6.05\times 10^{-3}$&  $\rm 3.14\times 10^{-3}$  & $\rm 1.27\times 10^{-3}$  &$\rm1.49\times 10^{-7}$\\
  & $\rm S/B$ & $\rm 3.93\times 10^{-2}$& $\rm 7.52\times 10^{-2}$  & $\rm 1.70\times 10^{-1}$ &$\rm 23.1$\\
 \hline
 \multirow{3}{2cm}{$\rm H\rightarrow \tau^+\tau^-$} &$\rm \sigma_{S}$ & $\rm4.01\times 10^{-5}$  &$\rm3.99\times 10^{-5}$& $\rm4.00\times 10^{-5}$ &$\rm3.41\times 10^{-6}$ \\
&$\rm \sigma_{B}$ & $\rm 5.42\times 10^{-4}$  &$\rm 2.86\times 10^{-4}$ & $\rm 6.44\times 10^{-5}$  &$\rm 2.23\times 10^{-7}$ \\
& $\rm S/B$ & $\rm 7.40\times 10^{-2}$& $\rm 1.40\times 10^{-1}$  & $\rm 6.21\times 10^{-1}$ &$\rm 15.3$\\
 \hline\hline
 \end{tabular}
 \end{center}
 \caption{\label{SBXsectionrrb}
Signal and total Background cross sections for $\rm e^- p \rightarrow \nu_e H b$ channel with different decay modes
after the application of Cut I-IV. The rejection factors and the b-tagging effects are taken into account in this table.}
\end{table}

\subsubsection{$\rm e^- p\rightarrow  \nu_e H b$ Channel with
$\rm H\rightarrow \gamma\gamma(b\bar{b},\tau^+ \tau^-$) Decay Modes}

We apply the similar method to $\rm e^- p\rightarrow  \nu_e H b$ production Channel.
However, due to the critical large backgrounds, we use much harder cuts instead:
For $\rm H\rightarrow \gamma\gamma$ decay mode,
Cut I still means the very basic cuts present in Eq.(\ref{basiccuts});
Cut II means Cut I plus $\rm |\eta^{jet}|<2.5GeV$, $\rm p_T^{b}>100GeV$, $\rm \Delta R (\gamma j)<4GeV$;
Cut III means Cut II plus invariance mass of diphoton pair belong to [$\rm m_h-3$, $\rm m_h+3$ ] GeV;
Cut IV means Cut III plus $\rm p_T^{\gamma}>150 GeV$, $\rm p_T^{b}>250 GeV$,
$\rm \Delta R(\gamma\gamma)<1.5GeV$. For $\rm H\rightarrow b\bar{b}$ decay mode, we use Cut II  to be
the basic cuts plus $\rm |\eta^{jet}|<2.5GeV$, $\rm \Delta R (\gamma j)<4GeV$,
and Cut IV to be the Cut III plus $\rm p_T^{b}>200GeV$.
For $\rm H\rightarrow \tau^+\tau^-$  decay mode, we use Cut II
to be the basic cuts plus $\rm |\eta^{jet}|<2.5GeV$, $\rm \Delta R (\gamma j)<4GeV$,
Cut IV to be the Cut III plus $\rm p_T^{b}>200GeV$,
$\rm p_T^{\ell}>125GeV$,$\rm \Delta R (\ell\ell)<1.5GeV$ . When jet fakes b, we replace the cuts for b to jets.
In Tab.\ref{SBXsectionrrb}, we display the signal and total background cross sections
after the application of Cut I-IV. Here in the table the rejection factors
and b-jet tagging efficiency are taken into account.

We see that in order to test the anomalous tqH coupling, the best choice of decay mode through $\rm e^- p \rightarrow \nu_e H b$ channel
is $\rm H\rightarrow b\bar{b}$. Though its cross section is much smaller than that of the one in
$\rm e^- p \rightarrow \nu_e \bar{t} \rightarrow \nu_e H \bar{q}$ channel with associated $\rm b\bar{b}$ decay mode,
its signal over background ratio is not small.
However, its cross section is small after the critical set of Cut IV which makes the detection a challenge.
By a simple fit we get
$\rm \sigma_{total} = 5.41 \times 10^{-9} + 7.43\times 10^{-9}\kappa_{tuH} + 8.00\times 10^{-6}\kappa^2_{tuH}$ [pb] for
$\rm H\rightarrow \gamma\gamma$.
$\rm \sigma_{total} = 1.46 \times 10^{-7} + 1.54\times 10^{-7}\kappa_{tuH} + 3.43\times 10^{-4}\kappa^2_{tuH}$ [pb] for $\rm H\rightarrow b\bar{b}$.
$\rm \sigma_{total} = 3.68\times 10^{-7} +1.86\times 10^{-7}\kappa_{tuH} +3.42\times 10^{-4}\kappa^2_{tuH}$ [pb] for $\rm H\rightarrow \tau^+\tau^-$.

\subsection{Data Analysis and Search Sensitivity}

\begin{figure}[hbtp]
\centering
\includegraphics[scale=0.6]{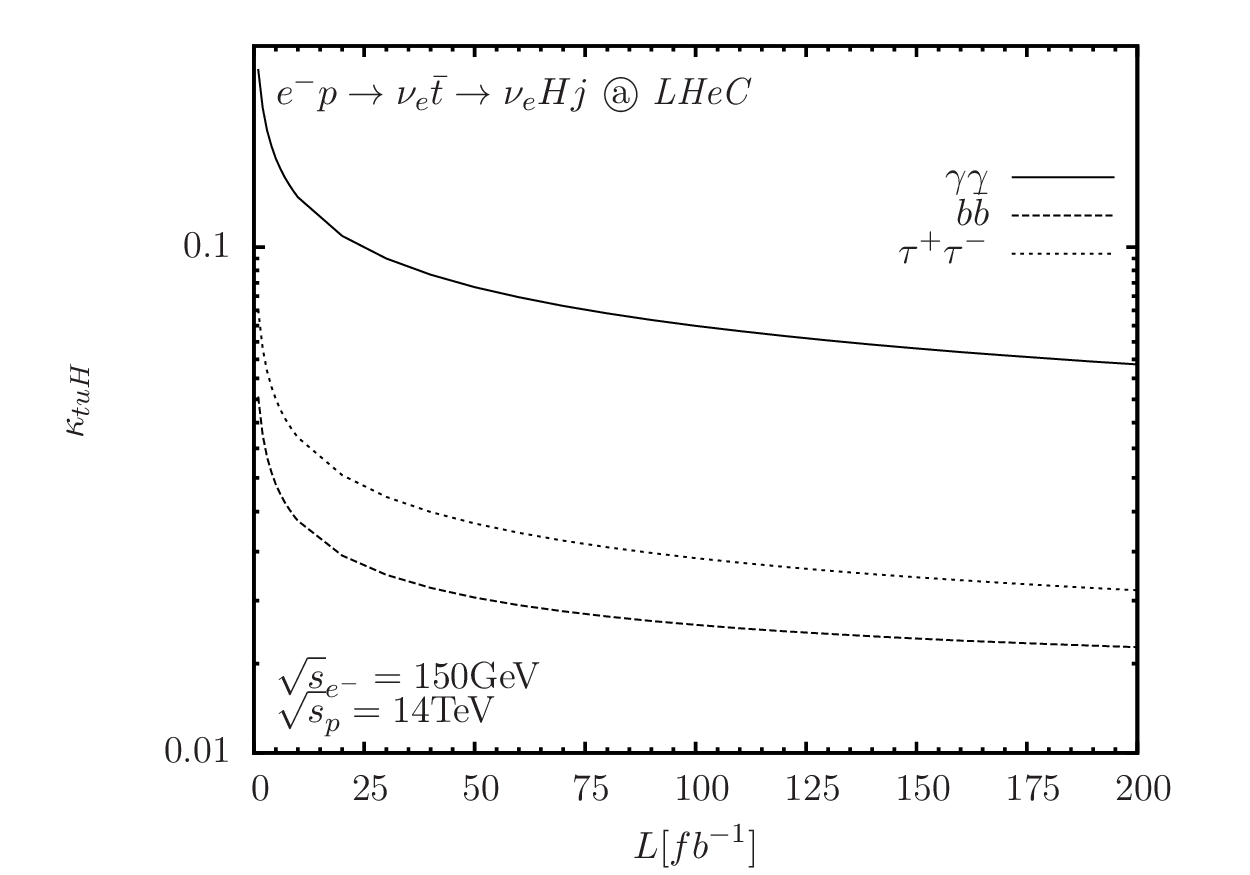}
\includegraphics[scale=0.6]{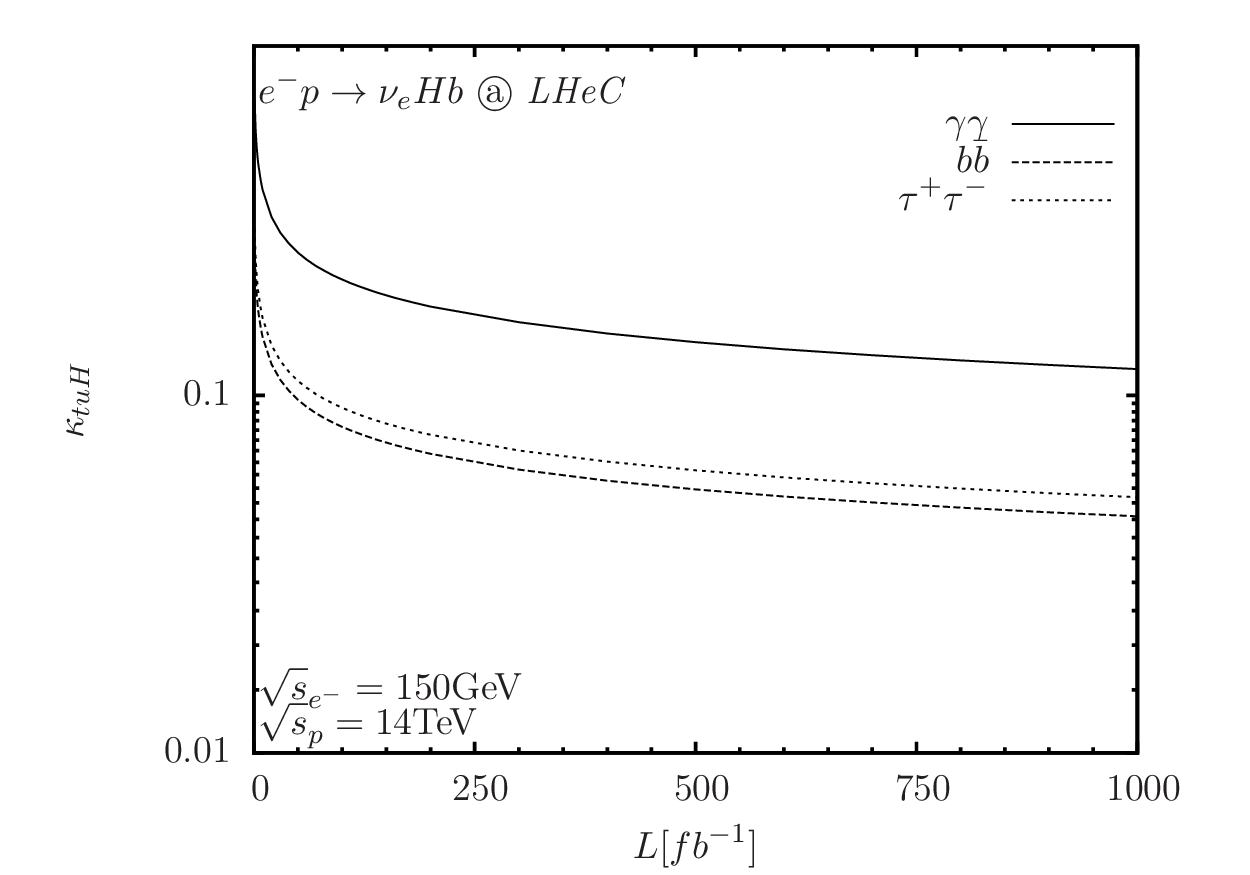}
\caption{\label{limit}
The contour plots in luminosity-$\rm \kappa_{tuH}$ plane for expected $95\%$ C.L. limits at 14 TeV LHeC.}
\end{figure}

We follow refs \cite{Anomaloustqr}\cite{tqr} exactly to obtain the sensitivity limits.
Typically, the limits are achieved by assuming the number of
observed events equal to the SM background prediction, $\rm N_{obs}=\sigma_{B} \times {\cal L} \times \epsilon$,
with $\cal L$ for a given integrated luminosity and $\epsilon$ the detection efficiency.
$\rm \sigma_{B}$ is the cross section of SM background prediction.
As can be seen, the SM background events can be less
or larger than 10 for different values of the luminosity.
We thus estimate the sensitivity limits on the anomalous $\rm tqH$ coupling
through both channels by using two different statistical analysis methods
depending on the number of observed events $\rm N_{obs}$.
For $\rm N_{obs} \leq 10$, we employ a Poisson distribution method.
In this case, the upper limits of number of events $\rm N_{up}$
at the 95$\%$ C.L. can be calculated from the formula
\begin{eqnarray}
\rm \Sigma^{N_{obs}}_{k=0} P_{Poisson} (N_{up};k)=1-CL .
\end{eqnarray}
Values for limits candidate $\rm N_{up}$ can be found in Ref.\cite{2014PDG}.
The expected $95\%$ C.L. limits on $\rm \kappa_{tqH}$
can then been calculated by the limits of the observed cross sections.
The integrated luminosity $\cal L$ will be taken as a running parameter.
For $\rm N_{obs} > 10$, a chi-square ($\chi^2$) analysis
is performed with the definition
\begin{eqnarray}
\rm  \chi^2 = (\frac{\sigma_{tot}-\sigma_{B}}{\sigma_{B}\delta})^2
\end{eqnarray}
where $\rm \sigma_{tot}$ is the cross section containing new physics effects
and $\rm \delta=\frac{1}{\sqrt{N}}$ is the statistical error
with $\rm N=\sigma_{B} \times {\cal L} \times \epsilon$.
The parameter sensitivity limits on anomalous $\rm tqH$ coupling
as a function of the integrated luminosity can then be obtained.

In Fig.\ref{limit}, we plot the contours of expected $95\%$ C.L. limits to
$\rm \kappa_{tuH}$ at 14 TeV LHeC with 150 GeV electron beam for
$\rm e^- p \rightarrow \nu_e \bar{t} \rightarrow \nu_e H \bar{q}$ [left panel]
and $\rm e^- p\rightarrow  \nu_e H b$ [right panel] channels respectively.
The solid curve, dotted curve and dashed curve are for $\rm H \rightarrow \gamma\gamma$,
$\rm H \rightarrow b\bar{b}$,
$\rm H \rightarrow \tau^+\tau^-$ decay modes respectively.
From Fig.\ref{limit}, we can see that the probed $\rm \kappa_{tuH}$ limits from
$\rm e^- p \rightarrow \nu_e \bar{t} \rightarrow \nu_e H \bar{q}$ channel
is much smaller than that from $\rm e^- p\rightarrow  \nu_e H b$ channel.
Typically, we get 0.0588, 0.0162, 0.0209 for $\rm \kappa_{tuh}$ by using
$\rm H \rightarrow \gamma\gamma$,
$\rm H \rightarrow b\bar{b}$, $\rm H \rightarrow \tau^+\tau^-$
decay modes respectively, which corresponds to the
branching ratio $\rm Br(t\rightarrow qH) = 0.177\%$, $\rm Br(t\rightarrow qH) = 0.0134\%$,
$\rm Br(t\rightarrow qH) = 0.0223\%$ at 14 TeV LHeC with 200 $\rm fb^{-1}$ luminosity for former channel,
and 0.177, 0.0701, 0.0776 for the latter, corresponding to the branching ratio
$\rm Br(t\rightarrow qH) = 1.604\%$, $\rm Br(t\rightarrow qH) = 0.252\%$,
$\rm Br(t\rightarrow qH) = 0.308\%$.
Thus, we apply higher luminosity for the latter channel, see, reaching to 1000 $\rm fb^{-1}$.
Then the research limits change to 0.118, 0.0468 and 0.0518 for $\rm \kappa_{tuH}$,
which corresponds to 0.713$\%$, 0.112$\%$, 0.137$\%$ for the branching ratio.
We can see that the LHeC sensitivity to the coupling $\rm \rm \kappa_{tuH}$ is much improved
by using $\rm e^- p \rightarrow \nu_e \bar{t} \rightarrow \nu_e H \bar{q}$ channel.
And for different decay modes, $\rm H \rightarrow b\bar{b}$ is best one for both channels.

\begin{table}[htp]
\begin{center}
\begin{tabular}{c c c c c c}
\hline\hline
 & \multicolumn{2}{c}{ $\rm e^- p\rightarrow \nu_e \bar{t}\rightarrow \nu_e H \bar{q}$ }
 & & \multicolumn{2}{c}{ $\rm e^- p\rightarrow  \nu_e H b$ } \\
                           & $\rm {\cal L} [fb^{-1}]$&$\rm Br(t\rightarrow qH)$ & &$\rm {\cal L} [fb^{-1}]$&$\rm Br(t\rightarrow qH)$\\
 \hline
  \multirow{2}{2cm}{$\rm H\rightarrow \gamma\gamma$} &10& 0.813$\%$  & &10&7.200$\%$  \\
 &200&0.177$\%$  &&200&1.604$\%$ \\
 \hline
  \multirow{2}{2cm}{$\rm H\rightarrow b\bar{b}$} &10& 0.0425$\%$  & &10&1.121$\%$  \\
 &200&0.0134$\%$  &&200&0.251$\%$ \\
  \hline
  \multirow{2}{2cm}{$\rm H\rightarrow \tau^+\tau^-$} &10& 0.0899$\%$  & &10&1.377$\%$  \\
 &200&0.0223$\%$ & &200&0.312$\%$ \\
 \hline\hline
 \end{tabular}
 \end{center}
 \caption{\label{summary}
Summary for the expected $95\%$ C.L. limits of $\rm Br(t\rightarrow qH)$
for $\rm e^- p\rightarrow \nu_e \bar{t}\rightarrow \nu_e H \bar{q}$ and $\rm e^- p\rightarrow  \nu_e H b$ channels
with $\rm H\rightarrow \gamma\gamma$, $\rm H\rightarrow b\bar{b}$, and $\rm H\rightarrow \tau^+\tau^-$ decay modes
at 14 TeV LHeC with 10(200) $\rm fb^{-1}$luminosity.}
\end{table}

In Tab.\ref{summary}, we give the $\rm Br(t\rightarrow qH)$
for different decay modes for both channels
at 14 TeV LHeC with 10(200) $\rm fb^{-1}$ luminosity respectively.
We see that the limits have improved by almost 4 times
when the luminosity increases from 10 to 200 $\rm fb^{-1}$.
When comparing different decay modes, $\rm H \rightarrow b\bar{b}$
is the best decay modes for both channels.
When we come to different channels,
$\rm e^- p \rightarrow \nu_e \bar{t} \rightarrow \nu_e H \bar{q}$ is much better than
$\rm e^- p\rightarrow  \nu_e H b$ channels by almost 10 times.
Finally, we use our best limits in $\rm H \rightarrow b\bar{b}$ decay modes for
$\rm e^- p \rightarrow \nu_e \bar{t} \rightarrow \nu_e H \bar{q}$ channel,
and we get 0.0134$\%$ for $\rm Br(t\rightarrow qH)$ as our result at 14 TeV LHeC with 200 $\rm fb^{-1}$ luminosity.

\section{Summary and Conclusion}

In this paper, we have investigated the anomalous flavor changing neutral current (FCNC) Yukawa interactions
between the top quark, the Higgs boson, and either an up or charm quark ($\rm tqH, q=u, c$).
We choose the channel $\rm e^- p\rightarrow \nu_e \bar{t} \rightarrow \nu_e H \bar{q}$
with $\rm q=u,c$ and the channel $\rm e^- p \rightarrow \nu_e H b$, where both channels are induced
by the charged current interaction through $\rm e^- p$ collision at the Large Hadron Electron Collider(LHeC).
We consider the $\rm H\rightarrow \gamma\gamma, b\bar{b}$ and $\tau^+\tau^-$ decay modes.
From the results, we can see that the flavor changing couplings $\rm \kappa_{tuH}$ can be probed to be minimal as 0.0162(0.0136) for the $95\%$ C.L. limits
in the $\rm e^- p \rightarrow \nu_e \bar{t} \rightarrow \nu_e H \bar{q}$ channel with $\rm H\rightarrow b \bar{b}$ decay mode,
which corresponds to the branching ratios $\rm Br(t\rightarrow qh)=1.34(0.947) \times 10^{-4}$ at 14 TeV LHeC with 200(3000) $\rm fb^{-1}$ luminosity.
From CMS and ATLAS Collaborations, we get the most stringent constraint of $\rm Br(t\rightarrow cH)< 0.56\%$,
$\rm Br(t\rightarrow uH)< 0.45\%$ at 95$\%$ confidence level (C.L.)\cite{FCNC_limit_tHq_CMS}.
Thus, we can see that our results shows a strong (above 30 times) improvement from experiments.
When comparing with the other phenomenological studies,
we can see that the LHeC sensitivity our results for $\rm Br(t\rightarrow qH)$ is smaller than the
sensitivity limits of LHC as $\rm Br(t\rightarrow qH)<5(2)\times 10^{-4}$ with
an integrated luminosity of 300(3000) $\rm fb^{-1}$ at $\rm \sqrt{s} = 14 TeV$\cite{HL_LHC_limit}.
Furthermore, our results are comparable with those of other studies, such as refs \cite{EFT_tqH_2}\cite{EFT_tqH_3}.
For example, ref \cite{EFT_tqH_2} obtains the sensitivity bound of about $0.1-0.3\%$ through different search channels
for an integrated luminosity of 100$\rm fb^{-1}$ at the $\rm \sqrt{s} = 13\ TeV$ LHC data.

\section*{Acknowledgments} \hspace{5mm}
Hao Sun acknowledges Fawzi Boudjema for his warm hospitality at  LAPTh.
Project supported by the National Natural Science Foundation of China
(Grant No. 11205070), by Shandong Province Natural Science Foundation
(Grant No. ZR2012AQ017), by the Fundamental Research Funds for the
Central Universities (Grant No. DUT15LK22) and
by China Scholarship Council (Grant CSC No. 201406065026) .

\end{document}